%% file: main2.tex
\documentclass{article}

\usepackage{subcaption}
\usepackage{booktabs} 
\usepackage{amssymb}
\usepackage{amsmath}
\usepackage{balance}
\usepackage{wrapfig}
\usepackage[pdftex]{graphicx}
\usepackage{tikz}
\usetikzlibrary{shapes}
\usepackage{pgfplots}
\usepackage{math-style}
\usepackage{enumitem}
\usepackage{stfloats}
\usepackage{placeins}
\usepackage{array}
\usepackage{ulem}
\usepackage{hyperref}
\usepackage{dsfont}
\usepackage{subcaption}
\usepackage{etoolbox}
\usepackage{mathabx}
\usepackage{placeins}
\usepackage{caption}

\renewcommand{\emph}[1]{\textit{#1}}

\usepackage[accepted]{icml2019}

\icmltitlerunning{Parallel Cut Pursuit}

\newcolumntype{C}{ >{\centering\arraybackslash} m{.16\textwidth}}

\newcommand*{\lightred}{red!30!white}

\newcommand*{\FV}{\upp{F}{\cV}}
\newcommand*{\dx}{\upp{d}{x}}
\newcommand*{\xiv}{\upp{\xi}{\cV}}

\newcommand*{\Eeq}{\upp{E_=}{x}}

\newcommand{\algorithmicdoinparallel}{\textbf{do in parallel}}
\makeatletter
\AtBeginEnvironment{algorithmic}{%
  \newcommand{\FORALLP}[2][default]{\ALC@it\algorithmicforall\ #2\ %
    \algorithmicdoinparallel\ALC@com{#1}\begin{ALC@for}}%
}
\makeatother

\tikzset{cross/.style={cross out, draw=red, minimum size=5mm, inner sep=0pt, outer sep=0pt, line width = .8mm, solid},
cross/.default={2mm}}

\begin{document}
\twocolumn[
\icmltitle{Parallel Cut Pursuit\\For Minimization of the Graph Total Variation}


\begin{icmlauthorlist}
\icmlauthor{Raguet Hugo}{blouah}
\icmlauthor{Landrieu Loic}{ign}
\end{icmlauthorlist}

\icmlaffiliation{blouah}{INSA Centre Val-de-Loire Universit\'e de Tours, LIFAT, France}
\icmlaffiliation{ign}{Univ. Paris-Est, IGN-ENSG, LaSTIG, STRUDEL, Saint-Mand\'e, France}
\icmlcorrespondingauthor{Raguet Hugo}{hugo.raguet@lilo.org}

\icmlkeywords{Graph, Total Variation, Cut Pursuit, Parallel}

\vskip 0.3in
]

\printAffiliationsAndNotice{} 

\begin{abstract}
We present a parallel version of the cut-pursuit algorithm for minimizing functionals involving the graph total variation. We show that the decomposition of the iterate into constant connected components, which is at the center of this method, allows for the seamless parallelization of the otherwise costly graph-cut based refinement stage. We demonstrate experimentally the efficiency of our method in a wide variety of settings, from simple denoising on huge graphs to more complex inverse problems with nondifferentiable penalties. We argue that our approach combines the efficiency of graph-cuts based optimizers with the versatility and ease of parallelization of traditional proximal splitting methods.
\end{abstract}

\section{Introduction}
In 2017, \cite{landrieu2017cut} introduced the cut-pursuit algorithm, a working-set algorithm for minimizing functionals involving the total variation structured by a graph $G=(V,E,w)$ , $w \in \bbR_+^E$ being  edge weights:
\begin{equation}
F: x \in \Omega^V \mapsto f(x) + \sum_{(u,v) \in E} w_{(u,v)} \Norm{x_u - x_v}~.
\end{equation}
where $x = (x_v)_{v \in V}\in \Omega^V$ is the variable of interest and $\Omega$ is some base space, typically $\bbR$ or $\bbR^n$, $n \in \mathbb{N}^*$.
The core idea of this approach is to exploit the \emph{coarseness} of the solution, \ie the fact that it can be decomposed into a small number of constant connected graph components of the graph compared to the total number of vertices.
This method was further refined by 
\citet{raguet2018cut}, allowing for the efficient regularization of a larger class of functionals. In particular, they 
dropped the convexity requirements in the optimality analysis, and extended the range of function $f$ to extended directionally differentiable functions with a possibly nondifferentiable part which is separable along the graph $G$. 

\begin{figure*}[ht!]
    \input{illustration}
    \caption{Illustration of the cut pursuit algorithm. The vertices of a graph $G$ are initially combined into a single component. This graph is recursively split into constant connected components. At each iteration, a reduced graph $\cG$ encoding the adjacency between constant components is computed. The values associated with each constant components can be computed using the smaller reduced graph. The vertex colors represent their values, from \emph{low} in blue to \emph{high} in red.}
    \label{fig:illustration}
\end{figure*}
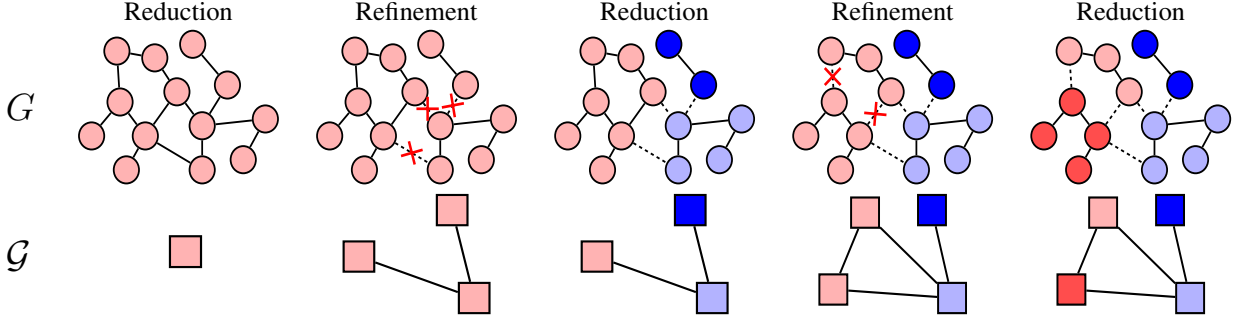

The cut-pursuit algorithm starts by associating all vertices to the same constant connected component and then repeats the two following steps: \emph{reduction} and \emph{refinement}. In the reduction step, the problem is solved under the constraint that all vertices in a same component have the same value. On many problems, when the number of constant connected components is smaller than the number of vertices in the graph, the reduced problem can be solved more efficiently than the original one, as it has fewer variables. In the refinement step, new ``degrees of freedom'' are added by splitting each constant connected component into smaller ones, based on a subproblem involving the directional derivatives of the functionals and which can be solved by graph cuts. This process is illustrated in \figref{fig:illustration}.

This scheme provably converges to a critical point in a finite number of steps; ensuring global optimality in the convex case. When the number of constant connected components of the solution is actually small, only a few iterations are needed for convergence. Succinctly, this approach allows to solve a wide range problems regularized by the graph total variation, in a little more than a few graph cuts.

In contrast, classical first-order proximal algorithms require many iterations, each of them updating all variables. 
Still, they attract considerable attention, notably because they can be preconditioned and easily parallelized; see for instance \cite{raguet2015preconditioning,mollenhof2018combinatorial,kumar2015convex,padilla2017dfs,barbero2017modular}.
On the other hand, the efficient use of graph cuts is reminiscent of the method of \cite{chambolle2009total}, which uses a parametric maximum flow formulation.
However, the latter can only be used to compute the \emph{proximity operator} of the graph total variation (also called ``graph total variation denoising'', or ``graph fused LASSO signal approximation''), that is $f$ restrained to a sum of square differences.
Not only can our method handle a much more general class of problems, but now that it can be parallelized, it combines the advantages of both proximal and graph cuts-based methods.
\section{Method}
%
\subsection{Reduced Problem}
We consider $\cV=\{U_1,\dotsc,U_{\Abs{\cV}}\}$ a partition of the vertices $V$ of  graph $G$. We define the \emph{reduced problem} as minimizing the objective functional $F$ under the constraint that the variable is constant over each component of the partition:
$$
\FV:\xi \in \Omega^\cV\mapsto F(\sum_{U \in \cV} \xi_U \otimes 1_U)~,
$$
where $\xi_U \otimes 1_U \in \Omega^V$ is a notation for $(\xi_U \otimes 1_U)_{v} = \xi_U$ if $v \in U$, $0$ otherwise.
In many problems, $\xi \mapsto f(\sum_{U \in \cV} \xi_U \otimes 1_U)$ has regularity and structure similar to $f$. But since the reduced problem can be rewritten as a problem of only $\Abs{\cV}$ variables, $\FV$ is easier to solve than $F$, typically with preconditioned first-order proximal splitting algorithms.
%
\subsection{Refinement}
%
The goal of the refinement step in the cut pursuit algorithm is to split the current partition $\cV$ such that the solution of the next reduced problem can decrease the objective functional $F$ as much as possible. This is done by considering first-order information through finding a steepest directional derivative of $F$ at the current iterate $x = \sum_{U \in \cV} \xi_U \otimes 1_U$, where $\xi$ is the solution of the last reduced problem: 
$$
\dx \in \argmin_{d \in \Omega^V} F'(x,d)~.
$$
Interestingly, to make the problem more tractable, the space of directions to consider can be restricted to 
a \emph{finite set} $D$.
\citet{landrieu2017cut} show that if $f$ is differentiable and $\Omega = \bbR$, $D = \{-1,1\}^V$ is sufficient to retain optimality at convergence.
\citet{raguet2018cut} extend this result to functions $f$ with a nondifferentiable part which is separable along the graph $G$, with $D=\{-1,0,1\}^V$. Furthermore, they provide heuristic direction sets $D^{(x)} = \bigtimes_{v \in V} D^{(x)}_v$ for multidimensional $\Omega$.

The search for a steepest descent direction restricted to a finite set $D$ defines a new partition by computing the cross partition between the current partition and the maximal constant connected components of $\dx$. This problem is a combinatorial optimization problem involving unary and binary terms:
\begin{align}
F'(x,d) = \sum_{v \in V} \delta(x,d_v) +\!\!\!\!\! \sum_{\uv \in \Eeq}\!\!\!\!\!  w_{u,v} \Norm{d_u - d_v}~,
\label{eq:cut}
\end{align}
where $\Eeq=\{\uv \in E \mid x_u = x_v\}$.
In the monodimensional case, this problem is equivalent to finding a \emph{minimum cut} in a convenient \emph{flow graph}; in the multidimensional case, it can be solved approximately \textit{via} a series of graph cuts.
\subsection{Parallelization}
In practice, the refinement step is often the computational bottleneck of the cut-pursuit algorithm. While there exists very efficient graph cut solvers, they require a lot of memory and are not well-suited to parallelization.

Now, since the only binary terms in \eqref{eq:cut} are for edges \emph{within} the constant components of $x$, the steepest descent problem is separable along the components of $\cV$:
\begin{align*}
F^\prime(x,d) & = \sum_{U \in \cV} F^\prime_U(x_U,d_U)~,
 \quad \text{with for all $U \in \cV$,} \\
F^\prime_U(x_U,d_U) & =\sum_{v \in U} F'(x_v,d_v)
 +\!\!\!\!\!\!\!\!\!\!\! \sum_{\uv \in \Eeq \cap U^2}\!\!\!\!\!\!\!\!\!  w_{u,v} {\Norm{d_u - d_v}}~,
\end{align*}
where we note $d_U=\{d_v\}_{v \in U}$.
Consequently, this problem can be decomposed into finding the steepest descent direction $\dx_U \in D^{(x)}_U = \bigtimes_{v \in U} D^{(x)}_v$ in each component $U$ independently.
This allows us to perform the graph cuts in parallel with only a slight adaptation of \emph{augmenting path} graph cut methods such as the one of Boykov and Kolmogorov (2004) \cite{boykov2004experimental}, and with no supplementary memory cost. 
%
\subsection{Balancing the Parallel Workload Distribution}
%
The above modifications already allows us to significantly speed up the cut-pursuit algorithm. However, thread utilization can still be improved. Indeed, when performing the split step in parallel along components, each component is assigned to a single thread, thus the computation is at least as long as the time required for splitting the hardest one, typically the largest.
If the partition is unbalanced, this can lead to most threads being idle while the last one finishes. Consider in particular the very first iteration, in which the partition only has one component.

A naive approach would be to cap the maximal number of vertices in each component of $\cV$ according to the problem size and the number of available threads; coherence with the graph structure could be enforced by greedily constructing each component with breadth-first search, until the component is exhausted, or until the cap is reached. However, such arbitrary \emph{balancing components} might not be beneficial for the reduced problem: in nonconvex settings, they might lead to bad local minima, and even in convex settings, they might not be consistent with the structure of the main problem and both slow down and reduce the accuracy of the solution.

For this reason, we advocate to decompose large components \emph{only for the split step}; then modifying the augmenting path algorithm accordingly to perform the graph cuts in parallel along these components. In a nutshell, this corresponds to isolating the balancing components by setting their border edge capacities to $0$ instead of $w_{u,v}\Norm{d_u - d_v}$ as they would be normally set to in the steepest direction problem. Unfortunately, the corresponding minimum cuts now solve only approximately the original problem.


As edges separating the balancing components are ignored by these parallel graph cuts, the question remains as how to define the new partition of a large component from the steepest directions found for its balancing components. Observe that all vertices of a component $U$ share the same set of candidate directions $\upp{D}{x}_U$, even in the multidimensional setting where it might change from one component to another. Consequently, the balancing components partitioning a given large component also share the same direction set. We can can thus refine $U$ from $\upp{d}{x}_U$ regardless of the presence of balancing components.

Note that this is not the only possible strategy. Depending on the problem at hand, it might still be relevant to apply the naive approach, or to compensate the discarded edge capacities by adjusting unary terms. However, the strategy presented here provided the most consistent improvement across our numerical experiments (alternative not shown).

We stress that solving only approximately the steepest direction problem invalidates the theoretical guarantees of the monodimensional setting, and might degrade the approximation further in the multidimensional setting. Although our experiments display this effect, they also show that such solutions remain satisfying at usual precision levels.

Let us finally note that there is still room for many improvements. For instance, finding rationales replacing the greedy breadth-first construction of the balancing components, or even randomizing it along iterations, might alleviate further the suboptimality. More importantly, in-depth study of the graph cuts complexity might suggest better parallel scheduling than a balancing only based on the number of involved vertices. This exploration is however beyond the scope of the present work, and left for future research.

\begin{algorithm}[tb]
   \caption{Parallel Cut Pursuit (PCP)}
   \label{alg:example}
\begin{algorithmic}
   \STATE {\bfseries Initialize:} $\cV=\Cur{V}$
   \REPEAT
  \STATE {}- - - - - \textit{reduced problem} - - - - -\\
   \STATE {\bfseries Find:} $\xiv$  stationary point of $F^{(\cV)}$\\
   $x \leftarrow \sum_{U \in \cV} \xiv_U \otimes 1_U$\\
   - - - - - \textit{parallel refinement} - - - - -\\
   \FORALLP{$U \in \cV$}
   \STATE {\bfseries Find:} $\dx_U\in D^{(x)}_U$ minimizing $F^\prime_U(x,d_U)$\\
   $\mathcal{U}$$\leftarrow$ max. constant connected components of $\dx_U$\\
   $\cV \leftarrow \cV \setminus \{U\} \cup \mathcal{U}$\\
   \ENDFOR\\
   \UNTIL{$\cV$ does not change}
\end{algorithmic}
\label{alg:pcp}
\end{algorithm}
%
\section{Numerical Experiments}
\begin{figure*}[ht!]
\input{xp2}
 \caption{Objective functional against the running time of the algorithms; optimal values are estimated by longest, high-precision runs.}
 \label{fig:xp}
\end{figure*}
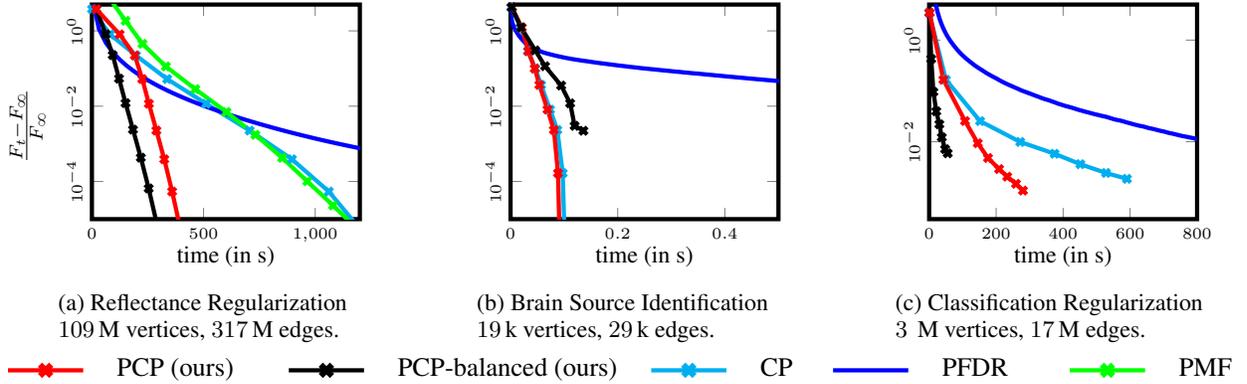
In this section, we show that our algorithm outperforms not only the highly specialized parametric maximum flow approach for the simpler problem of the proximity operator of the graph total variation, but also flexible preconditioned proximal algorithms on ill-conditioned problems involving several nondifferentiable regularizers beyond the total variation. All experiences are run on a $14$-cores, $28$-threads i9-7940X CPU \@ 3.10GHz with $64$ GB RAM.

Throughout this section, we compare our approach to the following state-of-the-art algorithms:
\begin{itemize}[leftmargin=2mm]
\setlength\itemsep{0mm}
\setlength{\itemindent}{0mm}
  \item \textbf{PMF:} the parametric max flow-based algorithm of \citet{chambolle2009total} for the proximity operator of the graph total variation;
  \item \textbf{PFDR:} the preconditioned Forward-Douglas--Rachford splitting algorithm of \citet{raguet2018note}, with all proximal steps parallelized;
  \item \textbf{CP:} the cut pursuit algorithm of \citet{raguet2018cut}, using PFDR to solve the reduced problem;
  \item \textbf{PCP:} our proposed parallelization of CP, without balancing the parallel workload distribution;
  \item \textbf{PCP-balanced:} PCP with balancing.
\end{itemize}
\noindent
\textbf{Point Cloud Reflectance Regularization:} We consider the problem of spatially regularizing LiDAR reflectance on a large point cloud acquired with a mobile scanning vehicle as described by \citet{paparoditis2012stereopolis}. The problem is defined as a graph-total variation denoising problem $f:x\mapsto\Norm{x-y}^2$ with $y$ the observed noisy reflectance. $G$ is the $5$-nearest-neighbors graph of the point cloud ($\Abs{V}=109\,412\,178$ and $\Abs{E}=$ $317\,096\,212$).

\noindent
\textbf{Brain Source Identification in Electroencephalography}
We consider the inverse problem of brain source identification in Electroencephalography. The brain of a patient is mapped to a triangular mesh with adjacency structure $G=(V,E)$ with $\Abs{V}=19\,626$ and $\Abs{E}=29\,439$.
A set of $N=91$ electrodes records the brain activity $y \in \bbR^N$ of the patient and the goal is to retrieve the activity on the mesh $G$. The relationship between the electrodes output and the brain activity is given by the \emph{lead-field} operator $\phi:\bbR^V \mapsto \bbR^N$. To model the regularity, sparsity and positivity of brain signals, we chose 
$$f:x\mapsto \frac12 \Norm{y - \phi x}^2 + \sum_{v \in V}\Pa{\lambda_v \Abs{x_v} + \iota_{\bbR_+}(x_v)}~,$$
with $\lambda_v$ the parameters of the weighted LASSO regularization and $\iota_{\bbR_+}$ the set indicator function of $\bbR_+$.

\noindent
\textbf{Point Cloud Classification Regularization} We consider the problem of spatially regularizing a noisy semantic labeling of a point cloud with class set $K$ \citep{hackel2017isprs}. $G$ is the $10$-nearest-neighbors graph of the point cloud, ($\Abs{V} =$ $3\,000\,111$ and $\Abs{E} =17\,206\,938$). A noisy classification $y \in \bbR^{V \times K}$ is obtained from a random forest classifier trained on handcrafted geometric features as described by \citet{guinard2017weakly}. Noting $\iota_{\Delta_K}$ the convex indicator of the standard simplex in dimension $\Abs{K}$, and $\KL{r}{s} =  \sum_{k \in K} r_k \log(r_k / s_k)$ the \emph{Kullback--Leibler divergence}, we choose 
$$
f\!:\!x \mapsto \KL{\beta u\!+\! (1\!-\!\beta) y_v}{\beta u\!+\!(1\!-\!\beta) x_v}+ \sum_{v \in V} \!\iota_{\Delta_K}(x_v),
$$
with $u = (1/\Abs{K})_{k \in K}$ the uniform discrete distribution and $\beta$ a smoothing constant (taken as $10^{-1}$).

We report performances in \figref{fig:xp}. All cut-pursuit approaches significantly outperform PFDR. Our parallelization scheme further increases this gap by a large margin for the large-scale point cloud regularization problems, even outperforming the specialized PMF. In contrast, parallelization brings virtually no gain for the medium-scale brain source identification problem. Then, balancing the parallel workload distribution of the split step is beneficial in terms of speed, in particular for the classification regularization problem, in which one of the components (the road in the middle of the scene) comprises half of the vertices. However, it can be seen that the balanced version stops at a suboptimal solution, which is nonetheless 
close to solutions found by the others (data not shown). Finally for the brain source identification problem, our balancing method is detrimental both in terms of speed and of accuracy. Our interpretation is that it misses the important sparsity structure of the problem and the strong correlation between the variables introduced by the lead-field operator.
\section*{Conclusion}
We introduce the first fully parallel graph-cut based approach for graph total variation minimization. It combines the advantages of traditional first-order proximal splitting algorithms and specialized graph cut-based methods. It significantly outperforms the state-of-the-art on many of the currently used total variation-regularization formulations.
We provide an implementation of our method, in C++ parallelized with OpenMP, with interfaces for GNU~Octave, Matlab and Python, at one of the authors GitHub repository \url{/1a7r0ch3/parallel-cut-pursuit}.
\FloatBarrier
\clearpage
\balance
{\small
\bibliographystyle{icml2019}

}

\clearpage
\balance
\section*{APPENDIX}
~
\begin{figure*}[!hb]
\input{illu}
 \caption{Illustration of the solutions given by the cut-pursuit algorithms; note that there is no discernable differences between the different versions. \Subref{fig:intensity2} noisy and regularized point cloud reflectance (detail of a much larger scene); \Subref{fig:eeg2} synthetic and recovered  ground truth brain activity; \Subref{fig:label2} noisy prediction and regularized point cloud classification. Images best seen on a monitor.}
 \label{fig:illu}
\end{figure*}
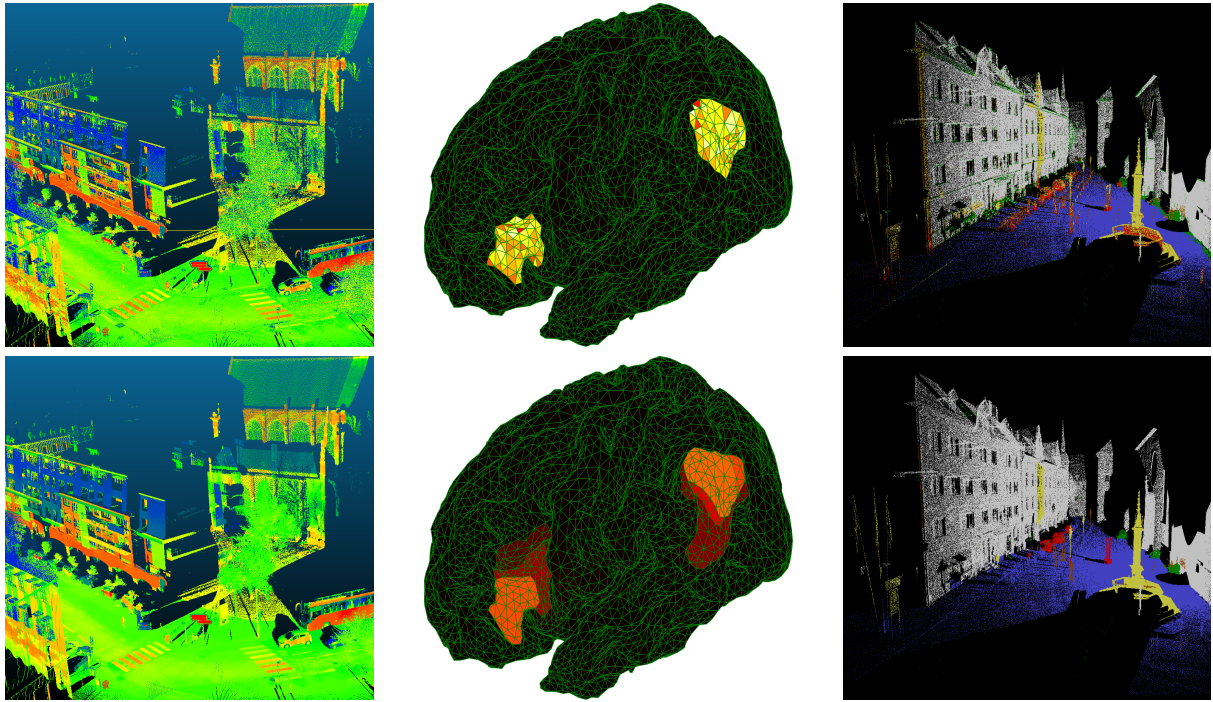
~
\end{document}

%% file: illustration.tex
\begin{tabular}{m{5mm}CCCCC}
  & Reduction & Refinement & Reduction & Refinement & Reduction
  \\
  \Large$G$
  &
    \resizebox{.16\textwidth}{.09\textheight}{
    \begin{tikzpicture}
      \input{./graph/main_graph_it0_sol.tikz}
    \end{tikzpicture}
    }
    &\resizebox{.16\textwidth}{.09\textheight}{
    \begin{tikzpicture}
                        \input{./graph/main_graph_it0_cut.tikz}
                      \end{tikzpicture}
                      }
                                 &
                                   \resizebox{.16\textwidth}{.09\textheight}{
                                   \begin{tikzpicture}
                                     \input{./graph/main_graph_it1_sol.tikz}
                                   \end{tikzpicture}
                                   }                & \resizebox{.16\textwidth}{.09\textheight}{ \begin{tikzpicture} \input{./graph/main_graph_it1_cut.tikz} \end{tikzpicture}                 }   &       \resizebox{.16\textwidth}{.09\textheight}{   \begin{tikzpicture}   \input{./graph/main_graph_it2_sol.tikz}          \end{tikzpicture}
                                                                  }
  \\
  \Large$\cG$
  &
    \resizebox{.12\textwidth}{.07\textheight}{
    \begin{tikzpicture}
      \tikzstyle{nod}=[rectangle, draw = black, fill = white, minimum size = 8mm, ultra thick]
      \node at (3.5, 3.5)     [nod, fill = \lightred]  (N2){};
      \node at (1.8143, 3.3714)     [nod, fill = none, draw = none]  (N2){};
      \node at (4.2500, 4.6000)     [nod, fill = none, draw = none]  (N3){};
      \node at (4.8250, 2.2750)     [nod, fill = none, draw = none]  (N4){};
    \end{tikzpicture}
    }
    &
      \resizebox{.12\textwidth}{.07\textheight}{
      \begin{tikzpicture}
        \input{./graph/reduced_graph_it0_cut.tikz}
      \end{tikzpicture}
      }
    &
      \resizebox{.12\textwidth}{.07\textheight}{
      \begin{tikzpicture}
        \input{./graph/reduced_graph_it1_sol.tikz}
      \end{tikzpicture}
      }
    &
      \resizebox{.12\textwidth}{.07\textheight}{
      \begin{tikzpicture}
        \input{./graph/reduced_graph_it2_cut.tikz}
      \end{tikzpicture}
      }
    &
      \resizebox{.12\textwidth}{.07\textheight}{
      \begin{tikzpicture}
        \input{./graph/reduced_graph_it2_sol.tikz}
      \end{tikzpicture}
      }
  \end{tabular}

%% file: xp2.tex
\addtolength{\belowcaptionskip}{+4mm}
    \begin{center}
\begin{tabular}{cccc}
    \begin{subfigure}[b]{.05\textwidth}
    \begin{tikzpicture}
    \node [draw = none] at (0,0) (n1) {};
    \node [draw = none, label={[label distance=0cm,text depth=0ex,rotate=90]left:$\frac{F_t-F_\infty}{F_\infty}$}] at (0.8,2.5) (n1) {};
    \node [draw = none] at (0,2.5) (n1) {};
    \end{tikzpicture}
    \caption*{~\\~}
    \end{subfigure}
    &
    \begin{subfigure}[b]{.30\textwidth}
        \begin{tikzpicture} \pgfplotstableread{./data/LIDAR_CP.dat}{\cp}
    \pgfplotstableread{./data/LIDAR_PCP.dat}{\pcp}
    \pgfplotstableread{./data/LIDAR_PCP2.dat}{\pcpb}
    \pgfplotstableread{./data/LIDAR_PFDR.dat}{\pfdr}
    \pgfplotstableread{./data/LIDAR_PMF.dat}{\pmf}
    \centering
    \begin{semilogyaxis}
    [width      = 1.0 \textwidth,
    xmin = 0.0, xmax = 1200,
    ymin = 0.00001, ymax = 5,
    xlabel={\small time (in s)},
    ,yticklabel style={font=\tiny,rotate=90},
    ,xticklabel style={font=\tiny,rotate=0},
    x label style={at={(axis description cs:0.5,+0.1)}},
    axis line style = ultra thick
    ]
    \addplot [cyan, ultra  thick, mark = x] table [x = {T}, y = {X}] {\cp};
    \addplot [blue, ultra thick , mark = none] table [x = {T}, y = {X}] {\pfdr};
    \addplot [black,ultra  thick, mark = x] table [x = {T}, y = {X}] {\pcpb};
    \addplot [red,ultra  thick, mark = x] table [x = {T}, y = {X}] {\pcp};
    \addplot [green,ultra  thick, mark = x] table [x = {T}, y = {X}] {\pmf};

      \pgfplotsset{every tick label/.append style={font=\tiny}}
     \end{semilogyaxis}
     \end{tikzpicture}
    \caption{Reflectance Regularization \\ $109$\,M vertices, $317$\,M edges.}
    \label{fig:reflectance}
    \end{subfigure}
    &
    \begin{subfigure}[b]{.30\textwidth}
    \begin{tikzpicture} \pgfplotstableread{./data/EEG_CP.dat}{\cp}
    \pgfplotstableread{./data/EEG_PFDR.dat}{\dr}
    \pgfplotstableread{./data/EEG_PCP.dat}{\pcp}
    \pgfplotstableread{./data/EEG_PCP2.dat}{\pcpb}
    \centering
    \begin{semilogyaxis}
    [width      = 1.0 \textwidth,
    xmin = 0.0, xmax = 0.5,
    ymin = 0.00001, ymax = 5,
    xlabel={\small time (in s)},
    ,yticklabel style={font=\tiny,rotate=90},
    ,xticklabel style={font=\tiny,rotate=0},
    x label style={at={(axis description cs:0.5,+0.1)}},
    axis line style = ultra thick
    ]
    \addplot [cyan, ultra  thick, mark = x] table [x = {T}, y = {X}] {\cp};
    \addplot [blue, ultra thick , mark = none] table [x = {T}, y = {X}] {\dr};
     \addplot [red,ultra  thick, mark = x] table [x = {T}, y = {X}] {\pcp};
      \addplot [black,ultra  thick, mark = x] table [x = {T}, y = {X}] {\pcpb};

      \pgfplotsset{every tick label/.append style={font=\tiny}}
     \end{semilogyaxis}
     \end{tikzpicture}
    \caption{Brain Source Identification \\ $19$\,k vertices, $29$\,k edges.}
    \label{fig:eeg}
    \end{subfigure}
    &
    \begin{subfigure}[b]{.30\textwidth}
        \begin{tikzpicture} \pgfplotstableread{./data/L3D_PCP2.dat}{\pcpb}
        \pgfplotstableread{./data/L3D_PCP.dat}{\pcp}
    \pgfplotstableread{./data/L3D_CP.dat}{\cp}
    \pgfplotstableread{./data/L3D_PFDR.dat}{\dr}
    \begin{semilogyaxis}
    [width      = 1.0 \textwidth,
    xmin = 0, xmax = 800,
    ymin = 0.0003, ymax = 5,
    xlabel={\small time (in s)},
    ,yticklabel style={font=\tiny,rotate=90},
    ,xticklabel style={font=\tiny,rotate=0},
    x label style={at={(axis description cs:0.5,+0.1)}},
    axis line style = ultra thick
    ]
    \addplot [cyan, ultra  thick, mark = x] table [x = {T}, y = {X}] {\cp};
    \addplot [blue, ultra thick , mark = none] table [x = {T}, y = {X}] {\dr};
    \addplot [black,ultra  thick, mark = x] table [x = {T}, y = {X}] {\pcpb};
    \addplot [red,ultra  thick, mark = x] table [x = {T}, y = {X}] {\pcp};
    
      \pgfplotsset{every tick label/.append style={font=\tiny}}
     \end{semilogyaxis}
     \end{tikzpicture}
    \caption{Classification Regularization \\ $3$\, M vertices, $17$\,M edges.}
    \label{fig:intensity}
    \end{subfigure}
\\
 \multicolumn{4}{c}{
\begin{tabular}{rlrlrlrlrl}
\begin{tikzpicture}[baseline={([yshift=-.5ex]current bounding box.center)}]
  \draw[red,ultra thick] (0,0) -- (1,0.0);
   \node[red, cross, scale = 0.3] at (0.5,0) (n1) {};
 \end{tikzpicture}
 &
 PCP (ours)\qquad
 \begin{tikzpicture}[baseline={([yshift=-.5ex]current bounding box.center)}]
  \draw[black,ultra thick] (0,0) -- (1,0.0);
   \node[cross, black, scale = 0.3] at (0.5,0) (n1) {};
 \end{tikzpicture}
 &
 PCP-balanced (ours)\qquad
 &
 \begin{tikzpicture}[baseline={([yshift=-.5ex]current bounding box.center)}]
  \draw[cyan,ultra thick,mark=*] (0,0)  -- (1,0.0);
  \node[cross,draw=cyan, scale = 0.3] at (0.5,0) (n1) {};
 \end{tikzpicture}
 &
 CP \qquad
 &
 \begin{tikzpicture}[baseline={([yshift=-.5ex]current bounding box.center)}]
  \draw[blue,ultra thick] (0,0) -- (1,0.0);
 \end{tikzpicture}
 &
 PFDR \qquad
  \begin{tikzpicture}[baseline={([yshift=-.5ex]current bounding box.center)}]
  \draw[green,ultra thick] (0,0) -- (1,0.0);
  \node[cross,draw=green, scale = 0.3] at (0.5,0) (n1) {};
 \end{tikzpicture}
 &
 PMF \qquad
\end{tabular}
}
\end{tabular}
     \end{center}
\addtolength{\belowcaptionskip}{-4mm}

%% file: illu.tex
\begin{tabular}{ccc}
    \begin{subfigure}[b]{.3\textwidth}
      \begin{tabular}{c}
    \includegraphics[width=.95\textwidth, height = .2\textheight]{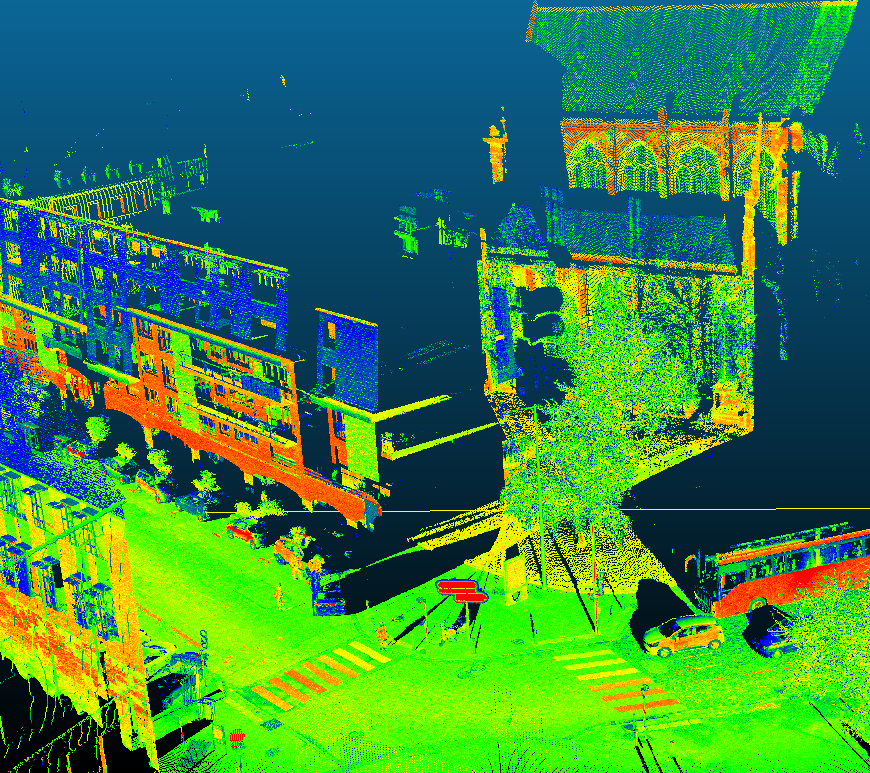}
    \\
    \includegraphics[width=.95\textwidth, height = .2\textheight]{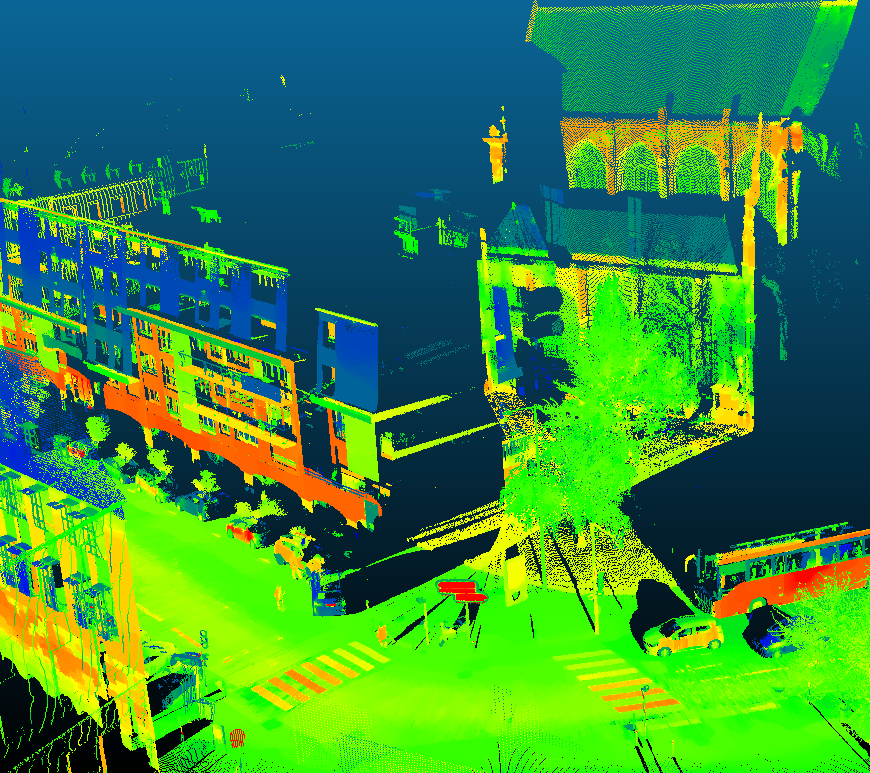}
    \end{tabular}
    \caption{Reflectance regularization}
    \label{fig:intensity2}
    \end{subfigure}
    &
    \begin{subfigure}[b]{.3\textwidth}
      \begin{tabular}{cc}
    \includegraphics[width=.95\textwidth, height = .2\textheight]{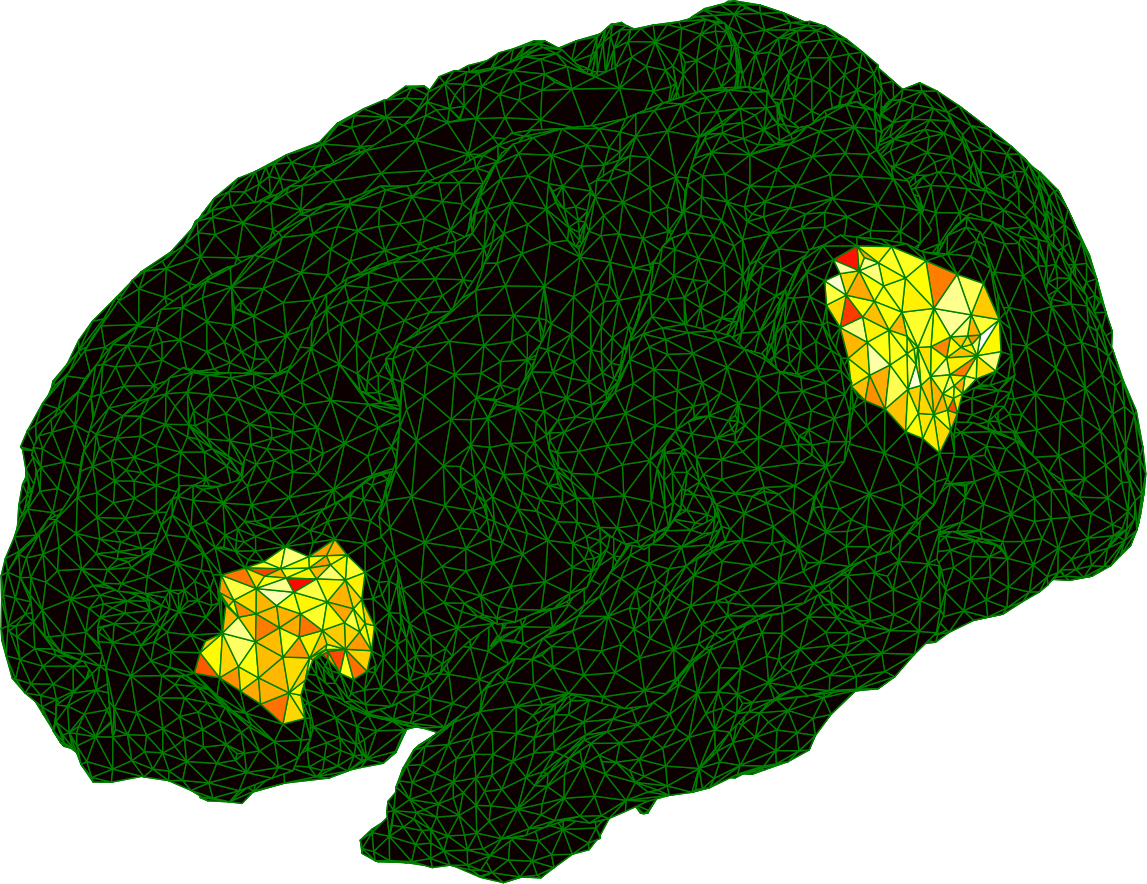}
    \\
    \includegraphics[width=.95\textwidth, height = .2\textheight]{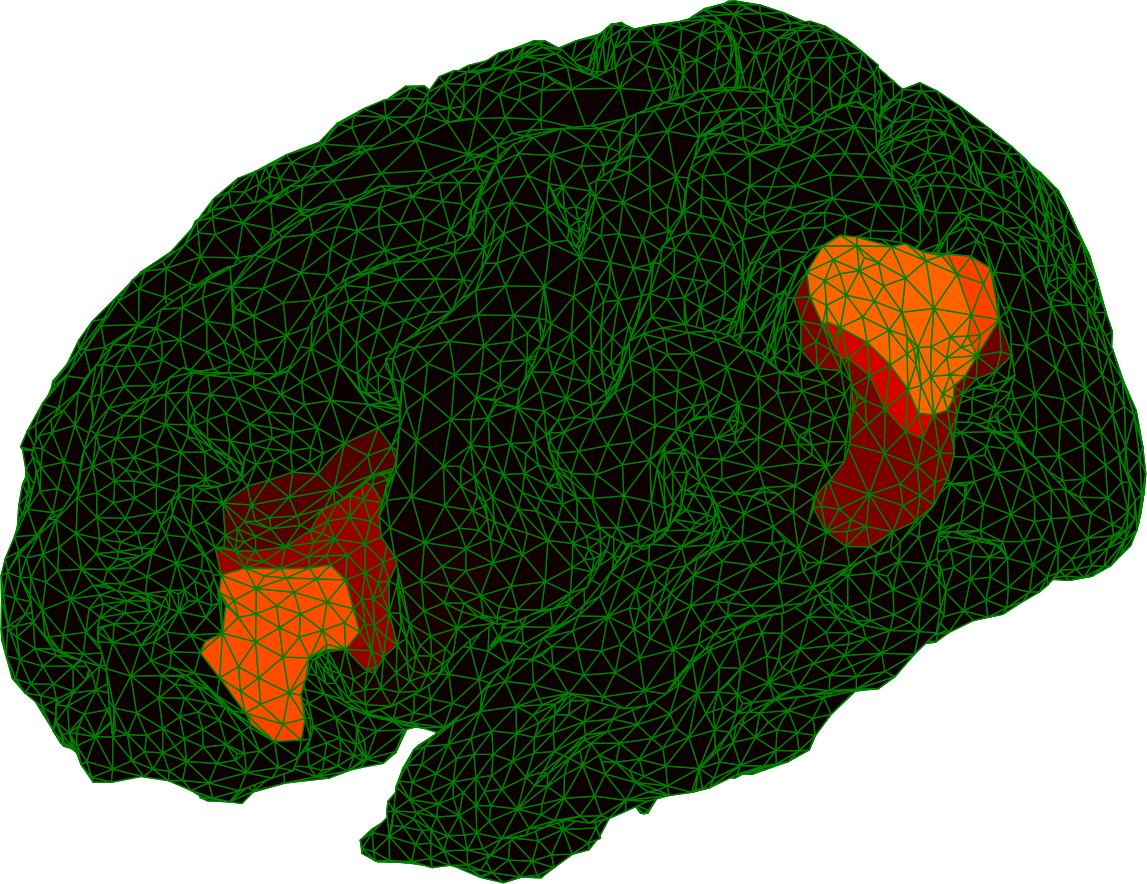}
    \end{tabular}
    \caption{Brain source identification}
    \label{fig:label2}
    \end{subfigure}
    &
    \begin{subfigure}[b]{.3\textwidth}
      \begin{tabular}{cc}
    \includegraphics[width=.95\textwidth, height = .2\textheight]{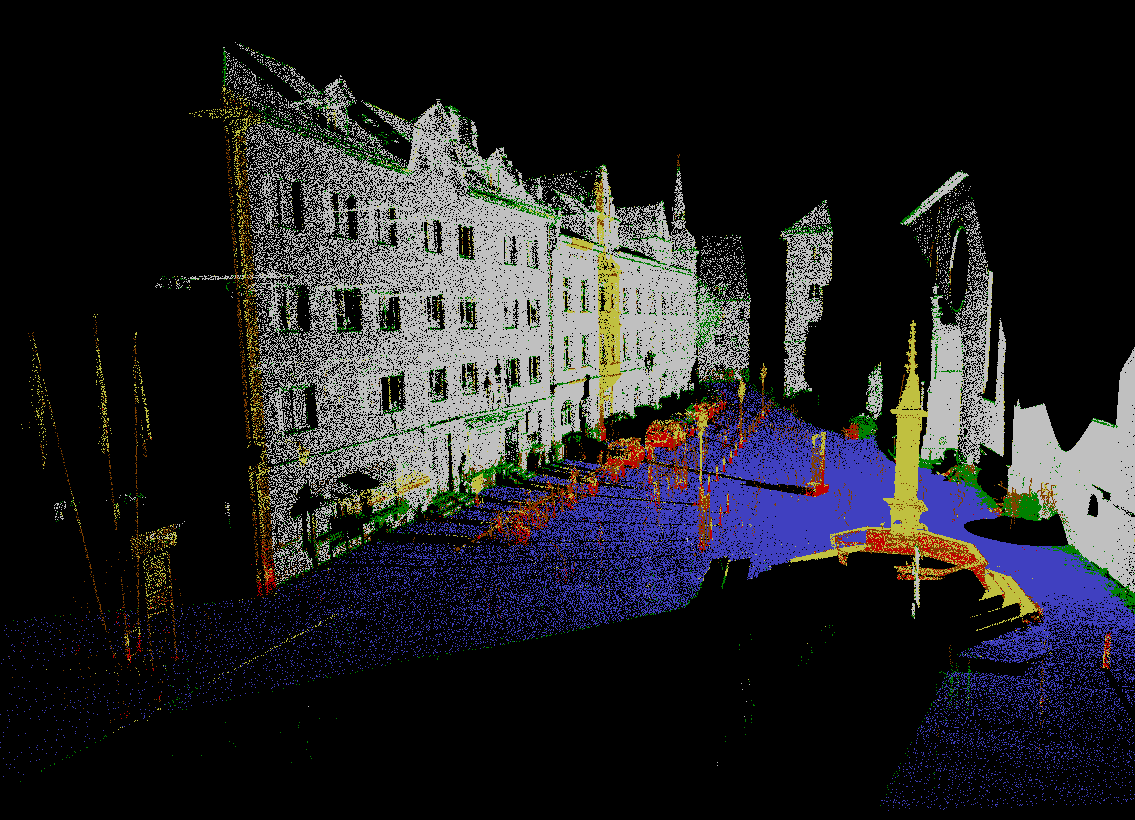}
    \\
    \includegraphics[width=.95\textwidth, height = .2\textheight]{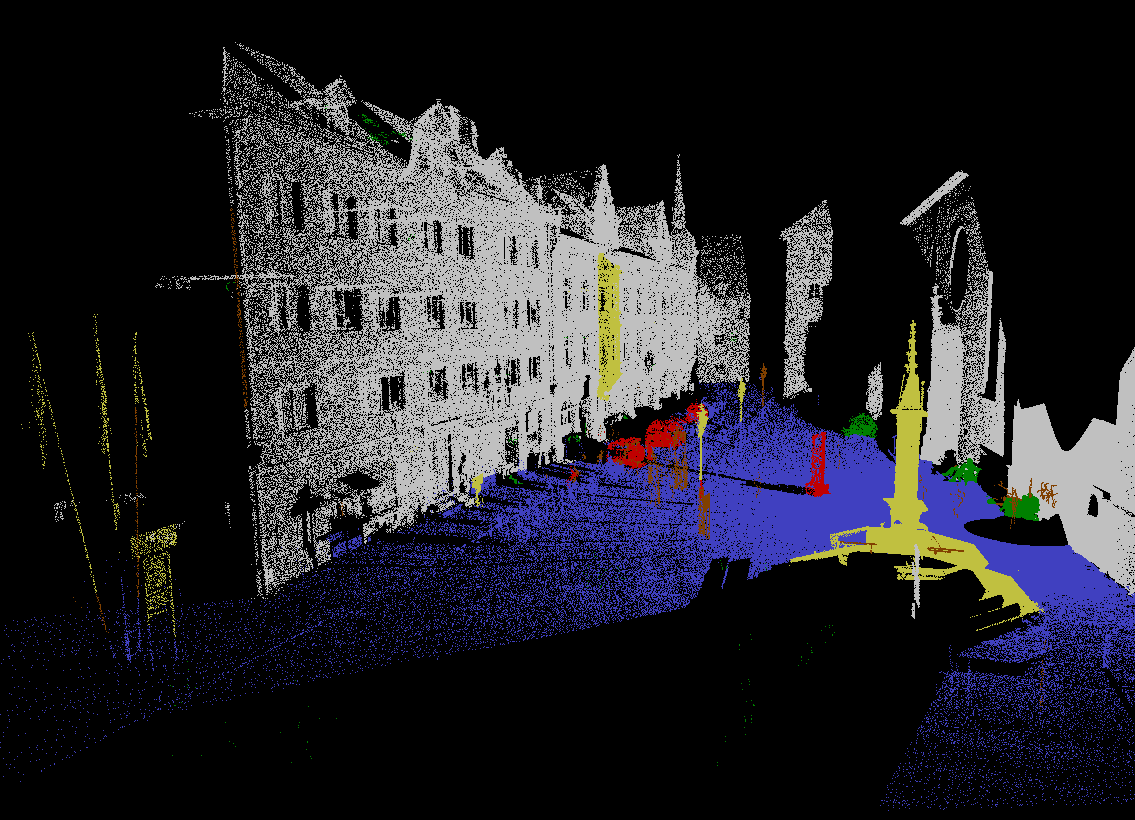}
    \end{tabular}
    \caption{Classification regularization}
    \label{fig:eeg2}
    \end{subfigure}
\end{tabular}